# Optimisation of SOUP-GAN and CSR-GAN for High Resolution MR Images Reconstruction

[1]Muneeba Rashid, [1]Hina Shakir, [1] Humaira Mehwish, [1] Asarim Amir, [2]Reema Qaiser Khan

1 Department of Software Engineering, Bahria University, Karachi, Pakistan

2 Department of Computer Science, Bahria University, Karachi, Pakistan

Contributing authors: muneebarashid55@gmail.com, hinashakir.bukc@bahria.edu.pk, humairamehwishkhan@gmail.com, asarimsesodia15@gmail.com, reemaqaiser.bukc@bahria.edu.pk

These authors contributed equally to this work
Corresponding author(s). E-mail(s): humairamehwishkhan@gmail.com

## Abstract

Magnetic Resonance (MR) imaging is a diagnostic tool used in modern medicine; however, its output can be affected by motion artefacts and may be limited by equipment. This research focuses on MRI image quality enhancement using two efficient Generative Adversarial Networks (GANs) models: SOUP-GAN and CSR-GAN. In both models, meaningful architectural modifications were introduced. The generator and discriminator of each were further deepened by adding convolutional layers and were enhanced in filter sizes as well. The LeakyReLU activation function was used to improve gradient flow, and hyperparameter tuning strategies were applied, including a reduced learning rate and an optimal batch size. Moreover, spectral normalisation was proposed to address mode collapse and improve training stability. The experiment shows that CSR-GAN has better performance in reconstructing the image with higher frequency details and reducing noise compared to other methods, with an optimised PSNR of 34.6 and SSIM of 0.89. However, SOUP-GAN performed the best in terms of delivering less noisy images with good structures, achieving a PSNR of 34.4 and SSIM of 0.83. The obtained results indicate that the proposed enhanced GAN model can be a useful tool for MR image quality improvement for subsequent better disease diagnostics.

## 1. Introduction

Magnetic Resonance (MR) imaging is one of the most critical diagnostic tools in clinical medicine, particularly in neurological and cardiovascular imaging. Although it is important, motion artefacts, noise interference, and low resolution are still obstacles to the accuracy and reliability of MR image scans. These problems hinder the quality of the diagnosis, reducing the possibility of healthcare specialists in detecting and diagnosing a disease correctly. These challenges have been addressed by traditional Image enhancement methods, which include filtering and interpolation. These techniques were considered sufficient for the image enhancement and reconstruction. However, with the passing time, they are declared not sufficient to capture fine details of the anatomy. Due to the advent of deep learning technologies, Generative Adversarial Networks (GANs) have become popular to improve the quality of MR images. [1]. Generative Adversarial Networks (GANs) have gained attention for their ability to reconstruct high-quality images from undersampled data, offering a promising solution for MR imaging[2].

Goodfellow et al. [3] introduced the GAN model that works on two components, known as the generator and discriminator. The generator reconstructs a synthetic image while the discriminator evaluates the image as real or fake. The outcome is an adversarial process where continuous upgrades are made, and this enables the generator to produce data that is more refined and indistinguishable from real data. This feature considers GANs as the most appropriate for medical imaging applications for precise and artefact-free image reconstructions.

This study conducts a comparative study of variants of GAN, i.e. SOUP. GAN and CSR-GAN, for reconstruction and enhancement of MR images. The SOUP-GAN generates smooth images to maintain the structures of the anatomy. In comparison, CSR-GAN aims at preserving the fine-grained details and, therefore, turns out to be highly effective in the enhancement of vital medical features. The optimisation of both models was carried out by adding more deep convolutional layers, making the filter sizes large, and enhancing the activation functions. The basic versions of both models were trained and evaluated on preprocessed MR image datasets. PSNR and Structural Similarity Index (SSIM) evaluation measures are used to measure the performance of these architectures. Improvement in the generator and discriminator controlled the training process, decreasing the artefacts and at the same time enhancing the performance of the two models in producing high-quality MR images.

The results of the experiment are discussed, compared and evaluated. The CSR-GAN with optimisation was better in artefact removal and structural preservation than SOUP-GAN. Conversely, SOUP-GAN stood out in producing globally consistent images. These results indicate that whereas CSR-GAN is superior in generating sharper images and preserving fine details, SOUP-GAN is still effective for broader image reconstruction tasks.

## 2. Related Work

In the field of medical image de-noising and enhancement, GANs proved to be valuable in MR image reconstruction. This is due to the prevalence of common imaging methods, such as CT imaging, which will normally discard important details that can be very important in clinical diagnosis [4]. They, however, perform very well in learning complex patterns in medical images and generating a high-quality image from a poor-quality scan. This ability has therefore seen GANs widely applied in tasks like noise removal, artefact correction and enhancing the resolution of MRI scans.

### 2.1. SOUP-GAN

The main GAN studied in the medical image enhancement is the Self-Organised Unsupervised Perceptual GAN (SOUP-GAN). Their designs emphasise image quality to the detriment of the necessary features for visual realism. Leadig et al. [5] helped build the primary fundamental ideas for the SOUP-GAN by proposing SRGAN, which in its turn used perceptual loss for creating high definition, realistic images. SOUP-GAN was developed to incorporate the perceptual learning into its work, and that helped them exquisitely balance between the preservation of Key data and provide desired outcomes.

The concept was further strengthened by Johnson et al. [6] by showing how perceptual loss from feature maps in pre-trained networks could improve image quality. Their study helped fine-tune the architecture of SOUP-GAN to handle fine textures. Similarly, Wang et Al [7] introduced ESRGAN and advanced GAN-based super-resolution with an improved version of the network. This approach influenced the deep architectural design of SOUP-GAN to allow better generation of texture and image quality. Haris et al. [8] proposed back projection networks that emphasize Iterative refining for resolution improvement. This iterative process was adopted in the generator of SOUP-GAN, enhancing its capability of recovering the missing details.

### 2.2. CSR-GAN

Another key GAN model is CSR-GAN that stands for Compressed Sensing Reconstruction GAN. It was specifically aimed at generating images based on compressed data. Hence, it is particularly intended towards MR image applications. It is capable of reconstructing high-quality images from limited data, which reduces the scan times. This is essential in medical imaging, where faster and correct scans takes towards effective and accurate diagnosis in healthcare.

Zhu et al. considered CSR-GAN for MR image reconstructions and compared it with traditional compressed sensing methods. They observed that CSR-GAN acquired an improved image quality and faster processing and became a game-changer in medical imaging. Chen et al. [9] examined various medical image reconstruction technologies. Therefore, they have noticed that CSR-GAN has the ability to handle highly compressed inputs without compromising the quality of the image. Mardani et al.[10] tried to enhance CSR-GAN by adding attention mechanisms and residual connections. This led to further superiority in fine detail preservation. These enhancements are essential for the medical applications when the critical anatomy of the human body must be represented accurately. Liu et al. [11] optimised the mathematical modelling of the training process of CSR-GAN to ensure that the learning process based on compressed data is stable and efficient.

SOUP-GAN is particularly good at generating aesthetically attractive images with smooth textures, whereas CSR GAN emphasises preserving fine details of the anatomy, crucial for accurate diagnosis. Both models incorporate perceptual and adversarial losses, balancing visual appeal and clinical accuracy. The continuous development of GAN-based models for medical imaging has led to significant improvements in MRI reconstruction. However, comparative studies remain limited, and standard evaluation methods still need refinement.

### 2.3. Significance of the Proposed Research Work

The contribution of this study is derived from the capability of enhancing the resolution and the quality of the computerised Cardiac MRI images, which in turn could enhance the correctness of medical diagnostics and is as follows:

I. Efficient and enhanced architectures of CSR-GAN and SOUP-GAN are proposed in an attempt to improve limited high-resolution MRI images without requiring costly equipment or time-consuming scans.
II. Creation of resource-efficient models to enhance the general performance of GAN models while at the same time making them portable and easily implementable for clinical purposes.

## 3. Materials and Methods

The proposed methodology outlines the comprehensive steps taken in enhancing MRI images using SOUP-GAN and CSR-GAN architectures and is demonstrated in Fig. 1. Following data collection and pre-processing, the baseline GAN models are built and trained. The architecture is enhanced, optimised, and retrained for the experimental data sets. The test images are reconstructed using optimised models, and the models' performance is evaluated.

### 3.1. Dataset Description

The MRI datasets used in this research consist of cardiac and neurological MRI scans from publicly available repositories. For the CSR-GAN, a CAD Cardiac MRI Dataset [12] comprising 2170 images was employed. This dataset consists of MRI images categorised into two main classes: "Normal" and "Sick."

In the case of SOUP-GAN, the Cardiac MRI data set of 2000 images available in Kaggle, known as the Sunnybrook Cardiac MRI Dataset [13] were used. The given dataset comprises a range of heart MRI images, which sample the suitable context for training the super-resolution GAN. The description of the data set is given in Table 1.

Table 1: Description of Datasets

|   | Dataset Name | Number of Images | Image Dimensions | Categories | Training / Test Split |
|---|---|---|---|---|---|
| 1 | Sunnybrook Cardiac MRI | 2000 | 256x256 | Multiple cardiac conditions | 70%/30% |
| 2 | CAD Cardiac MRI | 2170 | 200x200 | Normal, Sick | 70%/30% |

### 3.2. Pre-Processing of Datasets

Pre-processing steps were implemented to ensure compatibility with the GAN models. Resizing was performed to standardise image dimensions to 256×256 pixels for SOUP-GAN and 200×200 pixels for CSR-GAN. Normalisation was applied to pixel values, scaling them to a range of -1 to 1 to ensure stable model training. Thus, through pixel normalisation, the model can be able to easily compute gradients and weight updates in a better way. Grayscale conversion was applied to reduce the computational load of the given input data.

Moreover, data augmentation methods such as rotation, flipping, cropping, and zooming were employed to enhance dataset diversity, prevent overfitting and simulate real-world variations.

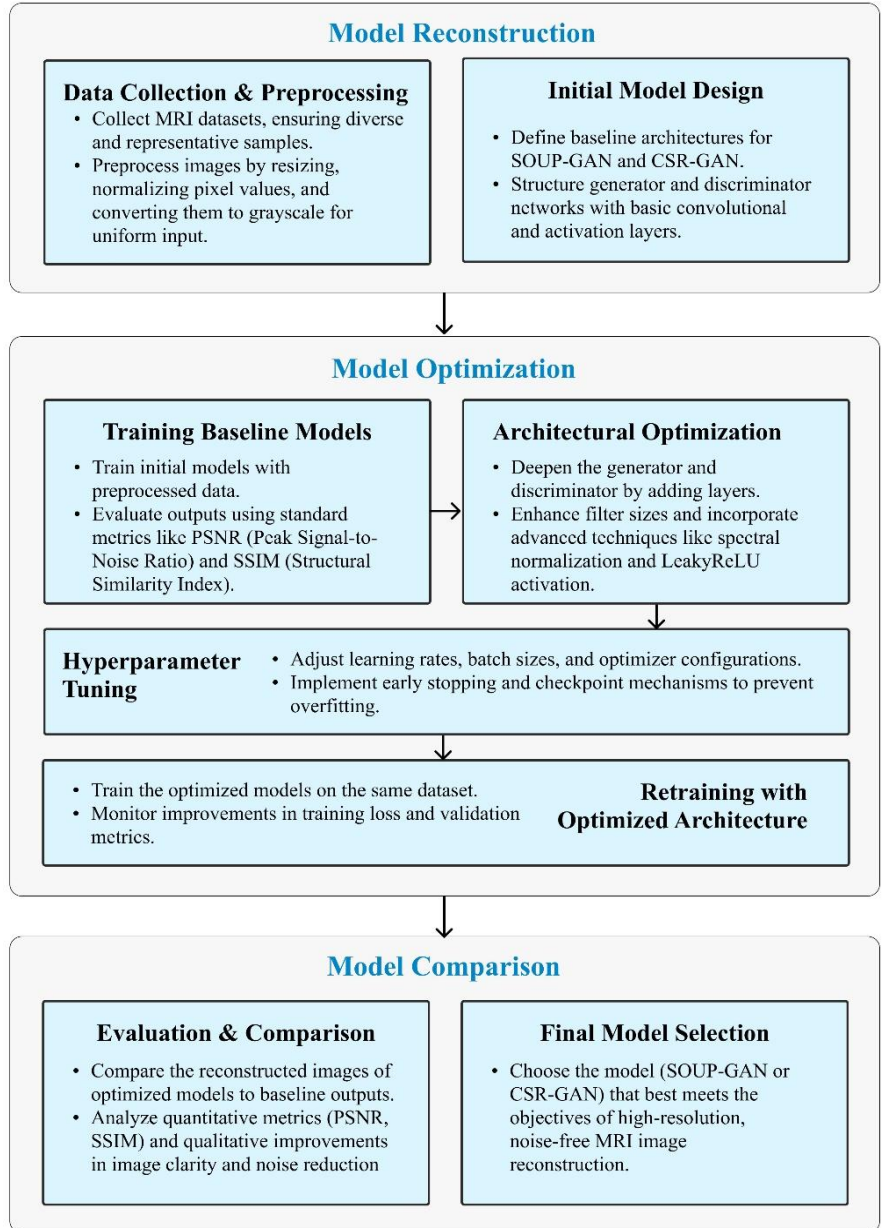

Fig. 1: Proposed Research Work Model

## 3.3. Enhancing SOUP-GAN Architecture

Standard SOUP-GAN's architecture contains a generator for producing high-resolution images from low-resolution inputs and a discriminator for distinguishing real images from generated ones. The key optimisation techniques included (i) the addition of convolutional deeper layers in the generator to extract complex features and (ii) replacing the ReLU Activation function with LeakyReLU to improve gradient flow.

Spectral normalisation was introduced to stabilise the discriminator and prevent mode collapse. Furthermore, an improved loss function through a combination of adversarial, perceptual, and pixel-based losses was used to enhance texture and overall image quality. The SOUP-GAN architecture before and after the proposed enhancement are shown in Fig. 2(a) and Fig. 2(b), respectively.

The architecture of SOUP-GAN in Figure 2(a) comprises a generator and a discriminator working in tandem. The process of the generator begins with a latent vector (size 100) as input, which is reshaped through a dense layer into a 3D tensor (64×64×256). This is followed by transposed convolution layers with LeakyReLU activation functions, progressively increasing the spatial dimensions from the original 64×64 pixels to 128×128 and finally ended up with a grayscale image size 256×256×1. The discriminator receives the input image (real or generated) by multiple successive convolutional layers, decreasing spatial dimensions from 256×256 to 32×32 while promoting feature depth to 256. It applies LeakyReLU activations in order to obtain efficient feature extraction, flattening the data to a dense layer that produces a single value for the classification of an image as real or fake. The combined model utilises binary cross-entropy as the loss function and the Adam optimiser to update the weights. The training process entails producing noise, generating fake images, and evaluating fake images against real ones. It also tracks losses of the generator and the discriminator. This process provides a base for adversarial learning. However, it does not support the enhancements for complex data and stability, which are considered in the optimised version.

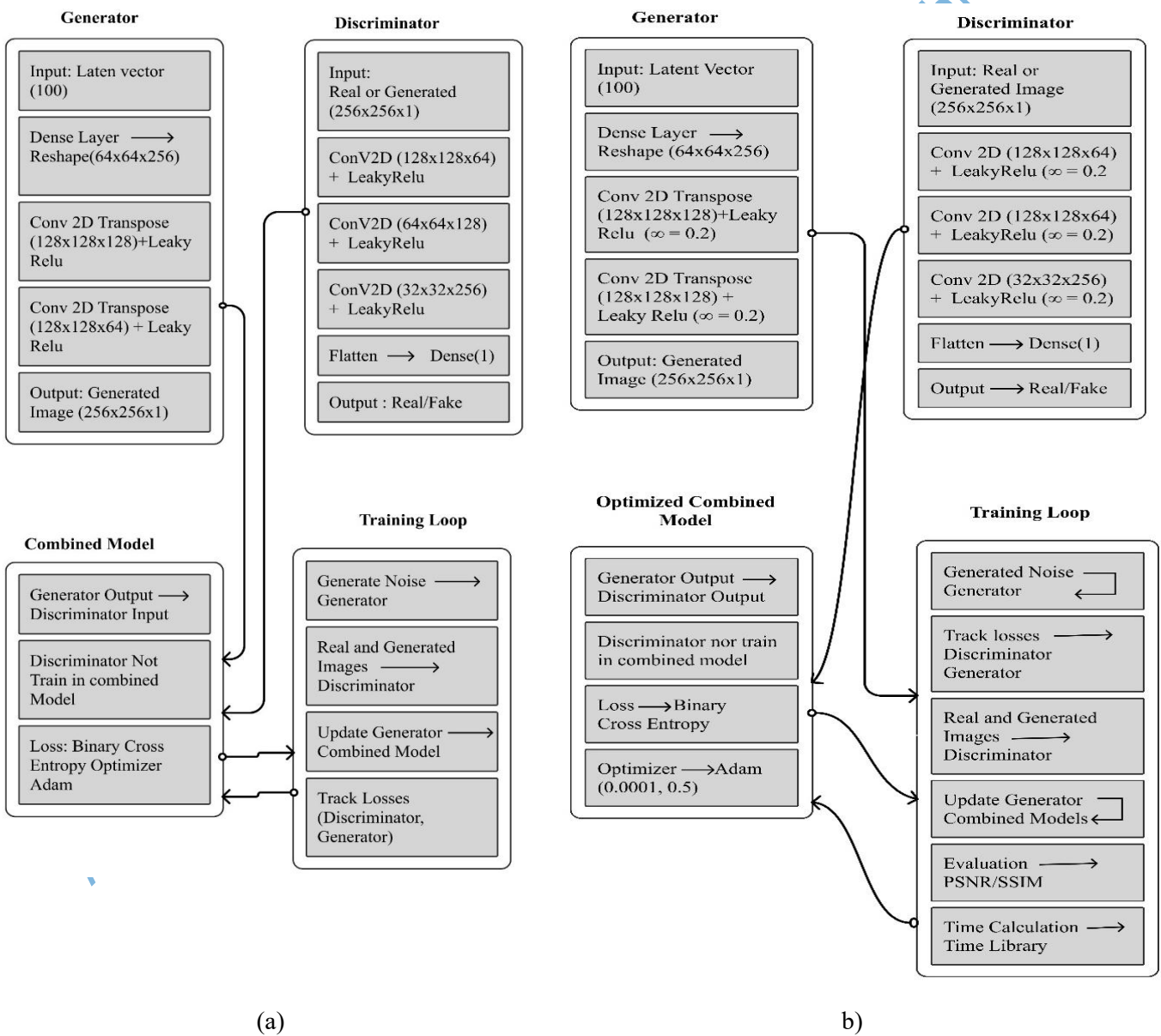

(a)            b)

Fig. 2 SOUP-GAN Architecture (a) Before Optimisation (b) After Optimisation

### 3.4. Enhancing CSR-GAN Architecture

Fig. 3(a) shows the CSR-GAN, where the process begins with the generator using two inputs as a noise vector (100 dimensions) and a condition vector (1 dimension). The randomness is added by the noise, and the input image start alining under certain conditions. A dense layer reformed these inputs into the 3D tensor with the dimensions 25x25x128. The successive transposed convolutional layers increase the scale of the tensor to the ultimate output of 200x200, having one channel, i.e. grey scale. The gradient stabilises, and feature extraction becomes enhanced by using the LeakyRELU activation function. The input image, whether real or generated, of size 200x200, is fed into the Discriminator with a condition vector. The combined input is then run through two convolutional layers for reducing the spatial dimensions, and the depth of the features increases from 100x100x64 to 50x50x128. The dense layer and flattening operation transform the extracted features into one probability value, and that is what defines whether the input is real or not.

The training process is accompanied by an adversarial procedure. The generator generates images using noise and conditions, while the discriminator discriminates these images comparing with the real images. Losses are calculated for both networks, and the discriminator is trained to better distinguish real from fake. While the generator is trained to produce images that can fool the discriminator. Loss tracking ensures that the models learn effectively and prevents any instability during training. This iterative process improves the generator and discriminator networks simultaneously, leading to better and higher-quality images. This baseline model proves the efficacy of the CSR-GAN's potential in high-resolution image reconstruction but lacks the advanced optimisation introduced later in the study.

The enhanced CSR-GAN architecture depicted in Fig. 3(b) consists of a conditional input generator that receives a noise vector as well as a label indicating the type of target image (normal/sick).   Next, the concept of residual blocks in the generator network was introduced to preserve the fine-grained details and enhance the feature learning. The advanced discriminator possesses knowledge of deeper convolutional layers and attention mechanisms to emphasise critical regions within the image. Also, enhanced loss functions include perceptual, adversarial, and reconstruction losses to balance realism and clinical relevance. The CSR-GAN architecture before and after the process of enhancement is illustrated in Fig. 3(a) & Fig. 3(b).

### 3.5. Hyper-parameter Optimisation

To maximise model performance, hyperparameter tuning has been conducted, with learning rates adjusted to 0.0001 for the generator and 0.0002 for the discriminator to ensure stable training. Batch sizes were optimised to 32 for SOUP-GAN and 64 for CSR-GAN, achieving a balance between computational efficiency and training quality. The hyperparameter tuning values are reported in Table. 2. Early stopping and model checkpoints were implemented to prevent overfitting and save the best model states. Early stopping facilitated halting training as soon as the losses remained constant for several epochs, while model check-pointing was performed at set intervals to resume training from a specific checkpoint in case of disruptions. Gradient clipping was applied to avoid unstable updates during backpropagation. Furthermore,

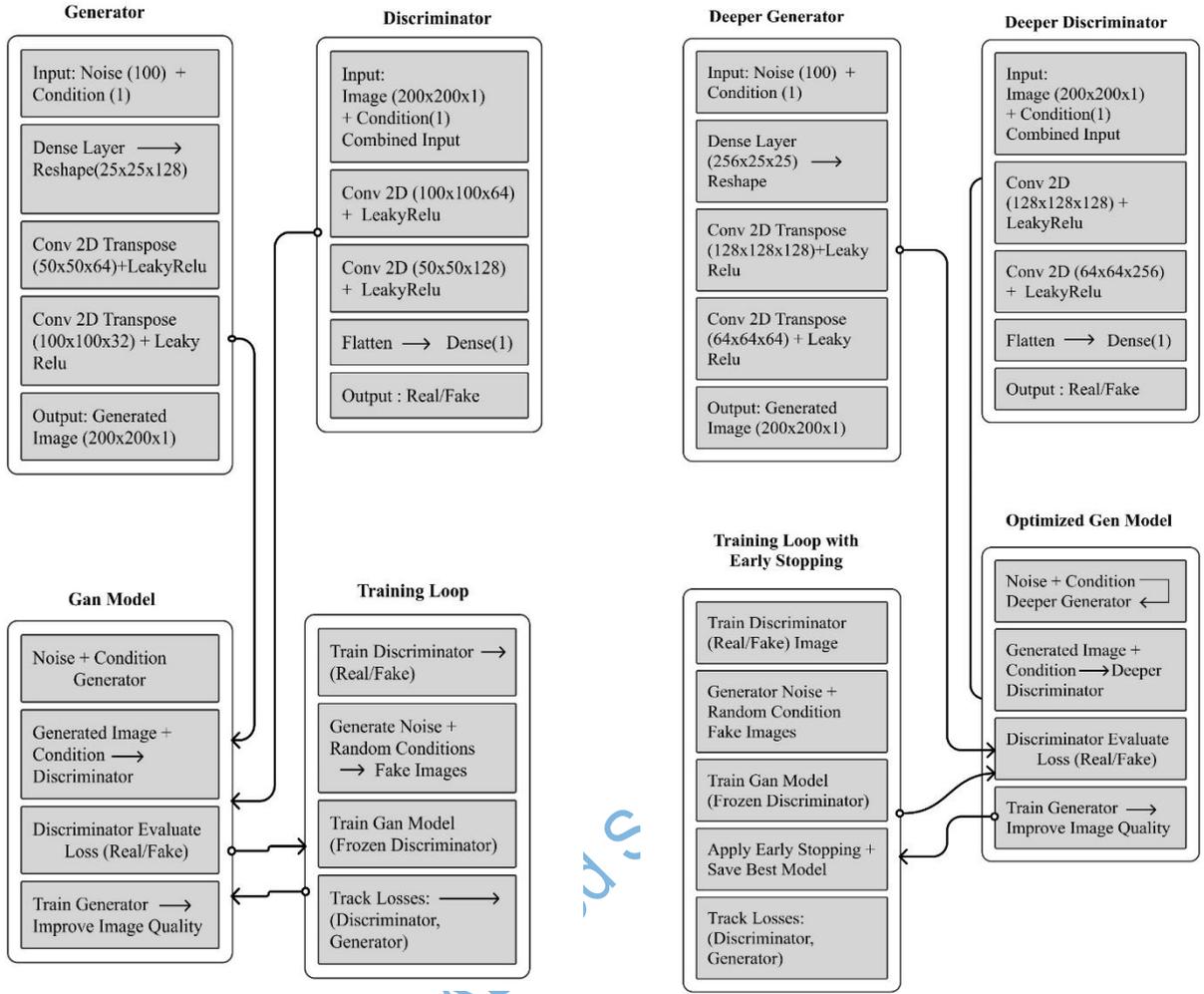

(a)                                                          (b)
Fig 3: CSR-GAN Architecture      (a) Before optimisation (b) After Optimization

spectral normalisation was used to ensure training stability by constraining the Lipschitz constant of the discriminator.

Table 2: Hyper-parameter Tuning

|   | Hyperparameter | SOUP-GAN Value | CSR-GAN Value |
|---|---|---|---|
| 1 | Learning Rate | 0.0001 | 0.0002 |
| 2 | Batch Size | 32.0 | 64.0 |
| 3 | Latent Dimension | 100.0 | 100.0 |

## 3.6. Evaluation Metrics

Two key evaluation metrics, which are well known to assess the quality of reconstructed high resolution image [14] using enhanced GAN models with respect to the original image, were used to assess the performance of enhanced GAN models.

The Peak Signal-to-Noise Ratio (PSNR) is the ratio of the maximum pixel value and the mean squared error occurring after image reconstruction and is given as follows:

$$\text{PNSR} = 10 \cdot \log_{10}\left(\frac{MAX^2}{MSE}\right) \qquad (1)$$

Where MAX is the maximum possible pixel value, and MSE is the Mean Squared Error between the original and generated images. A higher PSNR indicates less noise and better reconstruction.

Structural Similarity Index Measurement (SSIM) measures perceptual qualities based on luminance, contrast, and structure. It is defined as follows:

$$SSIM(x, y) = \frac{(2\mu_x \mu_y + C_1)(2\sigma_{xy} + C_2)}{(u_x^2 + u_y^2 + C_1)(\sigma_x^2 + \sigma_y^2 + C_2)} \qquad (2)$$

Where:

- $\mu_x$ and $\mu_y$ are the means of the original and generated images, respectively.
- $\sigma_x^2$ and $\sigma_y^2$ are the variances of the original and generated images.
- $\sigma_{xy}$ is the covariance between the two images.
- $C_1$ and $C_2$ are the constants to stabilise the division.

## 4. Results and Discussion

This section will showcase the quantitative and qualitative results gained from the execution of SOUP-GAN and CSR-GAN models for enhancing the MR images. PSNR and SSIM performance metrics are deliberated, exemplified by tables, training curves, and visual comparisons to emphasise enhancement achieved through optimisation.

### 4.1. Implementation Details

Models were executed using the latest industrial frameworks. TensorFlow 2.0 and PyTorch were used for building, training and evaluating models. Numpy, OpenCV, Matplotlib, as well as data support for Scikit-learn, facilitated data processing and visualisation. The training process was performed on the NVIDIA Tesla V100 GPU card with 32GB VRAM, which ensures fast and efficient computation. This is achieved by the integration of recent advancements in GAN architectures with advanced optimisation techniques and state-of-the-art tools. Hence, the quality of reconstructed MR images were enhance that facilitates more reliable medical diagnosis and clinical research.

### 4.2. Quantitative Results

In quantitative analysis, PSNR and SSIM results were achieved before and after model optimisation. Table. 3 lists the comparison of the results of SOUP-GAN and CSR-GAN. Furthermore, the results are depicted in Fig. 4(a) & (b) for PSNR and SSIM metric, respectively. The optimised models made considerable progress, and CSR-GAN outperformed in terms of noise reduction and structure preservation.

Table 3: Comparison of PSNR and SSIM

| Model | PSNR (dB) (Before) | PSNR (dB) (After) | SSIM (Before) | SSIM (After) |
|---|---|---|---|---|
| SOUP-GAN | 28.7 dB | 34.4 dB | 0.67 | 0.83 |
| CSR-GAN | 31.2 dB | 34.6 dB | 0.74 | 0.89 |

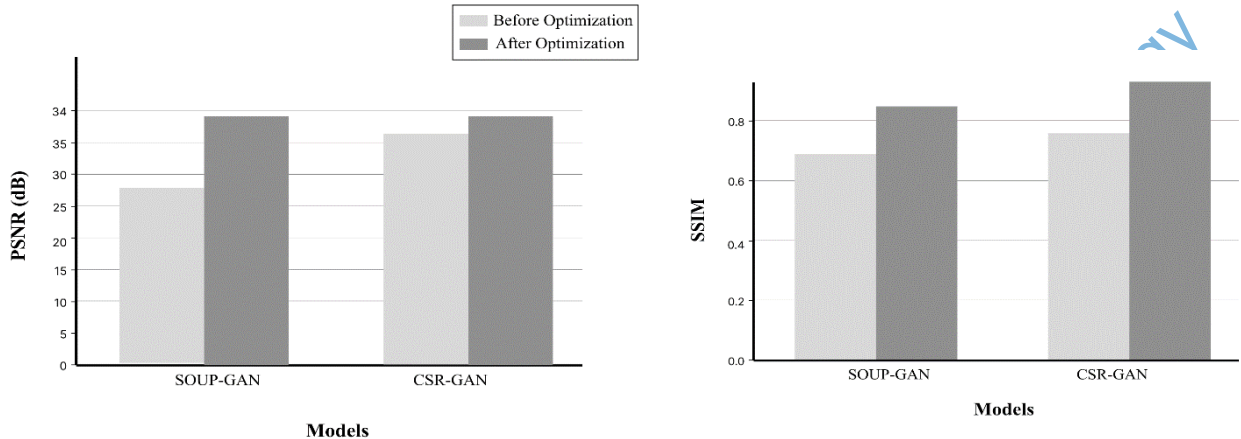

Fig. 4: Before and After Optimization     (a) PSNR Comparison (b) SSIM Comparison

The results showed significant improvements for the optimised models, with CSR-GAN performing better than other models.

### 4.3. Training Time Comparisons

The training time of the models varied considerably before and after the optimisations because of architectural improvements and hyper-parameter tuning, as shown in Tab. 4. Before optimisation, the time for each epoch in training was comparatively higher as the models faced inefficiencies in gradient flow and stability. For instance, SOUP-GAN took about 2.5 minutes for each epoch, while CSR-GAN required approximately 3 minutes for every epoch, indicating limitations of the original architectures.

### 4.4. Qualitative Analysis

Visual comparisons were performed by comparing the generated MR images before and after optimisation. Figs. 5 and 6 depict typical samples from both models. The results clearly present the quality of the image improved step by step, from the original, noisy and blurry MR images to the reconstructed images from the GAN models and finally to the optimised versions. The original images were full of noise, motion artefacts, and lacked clarity, making them less useful for medical diagnosis. Initial GAN models improved these images, but the result still contained some noise and artefacts, and the finer details were not preserved very well. These issues were dealt with efficiently by the optimised architectures, namely SOUP-GAN and CSR-GAN. SOUP-GAN managed to produce smoother pictures while maintaining the integrity of the original structure intact, making it great for cases where maintaining the original layout of the image is considered important. However, CSR-GAN was far superior in the preservation of small features and details with reduced noise, providing the image resolution necessary for the demonstration of fine anatomical structures.

After optimisation, when the training time for one epoch was kept constant, the proposed model achieved significantly better image quality in the same time frame, demonstrating an efficient trade-off between computational cost and output quality. The optimised model yielded higher PSNR and SSIM values, providing higher quality results without increasing the computational cost. Hence, demonstrates efficient optimisation. The dataset size stays the same for both before and after the optimisation, but has a direct effect on training time per epoch because of batch processing and data handling throughout the overall training.

These results evidently show a huge improvement in image quality after optimisation of the GAN architectures. In Fig. 5, the improved SOUP-GAN obtained the SSIM of 83% as compared to an SSIM of 67% for the Sunnybrook Cardiac MR images dataset. The improvement in SSIM indicates that it generates images that are smoother and have less noise while maintaining the overall structure. In medical imaging, clear and smoother images are required to make a proper diagnosis. Similarly, for the CAD MR images dataset, the CSR GAN observes the SSIM value 74% before the optimisation process, which further improved to 89% with improved SOUP-GAN, proving to be a significant enhancement of 15%.

This proposed research addresses these gaps by conducting a comprehensive comparison of SOUP-GAN and CSR-GAN regarding their use in MR images enhancement. Advanced Optimization techniques like deeper network architectures, refined loss functions, and better training procedures are applied to both models. The ultimate goal is to advance medical imaging capabilities, enabling more accurate diagnoses and better patient outcomes through state-of-the-art image enhancement methods.

Table 4: Comparison of Training Time

| Metric | SOUP-GAN (Before Optimisation) | SOUP-GAN (After Optimisation) | CSR-GAN (Before Optimisation) | CSR-GAN (After Optimisation) |
|---|---|---|---|---|
| **Training Time Per Epoch** | ~2.5 minutes | ~2.5 minutes | ~3 minutes | ~3 minutes |
| **Number of Epochs** | 40 | 40 | 40 | 40 |
| **Total Training Time** | ~100 minutes | ~100 minutes | ~120 minutes | ~120 minutes |
| **Optimization Impact** | Inefficient gradient flow and stability | Improved architectures and better gradient flow | Inefficient gradient flow and stability | Improved architectures and better gradient flow |

### 4.5. Comparison with the state of the art

In the literature, Waqar Ahmad [15] proposed a robust GAN-based SR framework that addresses common limitations of previous methods, which include single-scale feature extraction and single-step upscaling. The proposed framework resulted in stronger enhancement and achieved an average PSNR of 30 dB and SSIM of 0.865. The achieved results improved super-resolution but lacked fine detail preservation.

Zhu et al. [16] proposed CSR GAN with adversarial learning for MR images reconstruction on perceptual optimisation with SSIM OF 0.82. The proposed research resulted in better noise reduction and enhanced MR images reconstruction, and achieved a PSNR of 33.2 dB and SSIM of 0.85.

The authors [17] enhanced SOUP-GAN further by modifying its architecture, introducing additional convolutional layers, optimising hyperparameters, and refining the loss functions, but lacked a fine-grained structure. The proposed framework resulted in a stronger enhancement of the SSIM of 32 dB and an accuracy of 82%.

The presented research goes beyond by comparing two GAN architectures, SOUP-GAN and CSR-GAN evaluating their performance in different enhancement tasks, and optimising their structures to improve reconstruction quality. The proposed research extends Wang et al. [18]'s work, who originally introduced CSR-GAN to reconstruct MR images from limited measurements. In the current study, the proposed model [18] has been improved. It results in noise reductions and has higher PSNR due to the inclusion of more layers, spectral normalisation, and improved activation functions. This proposed study is distinct in aiming to optimise the existing models of CSR-GAN and SOUP- GAN rather than proposing a new model, ensuring efficiency in MR images enhancement tasks. In Table 5, a quantitative comparison of the presented research with the state of the art is reported.

      The presented research work compares two well-established GAN architectures (SOUP-GAN and CSR-GAN) rather than developing an entirely new GAN model. Performance-wise, the proposed optimised CSR-GAN achieved a PSNR of 34.6 dB and an SSIM of 0.89, outperforming the baseline versions in previous studies. The optimised model addressed MR images motion artefacts and noise reduction more effectively than some of the past work by implementing advanced regularisation techniques, spectral normalisation, and modified activation functions. My research aimed to balance computational efficiency and model accuracy, unlike some prior studies that focused solely on model complexity without considering resource constraints.

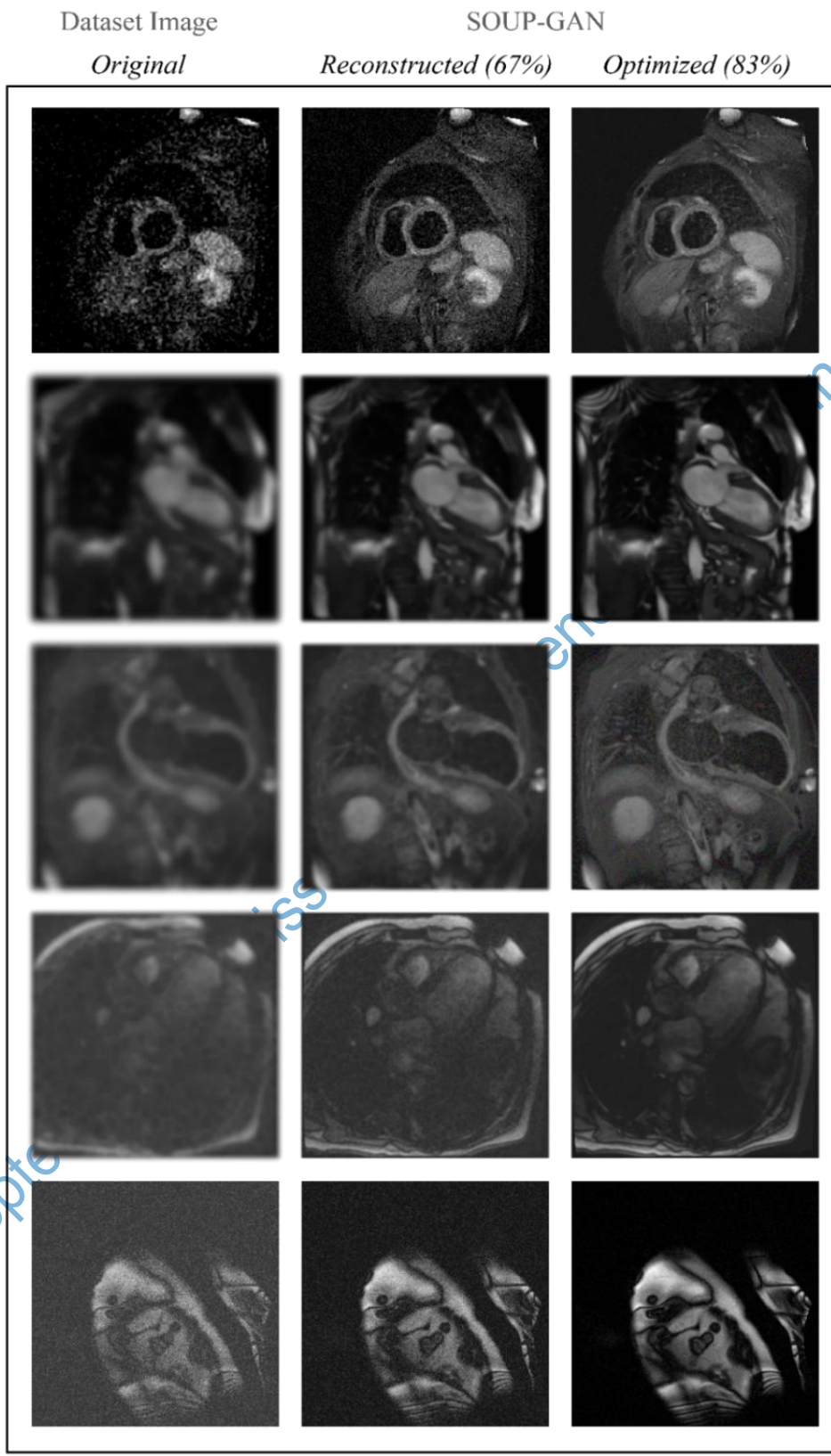

Fig. 5: SOUP-GAN Testing Results

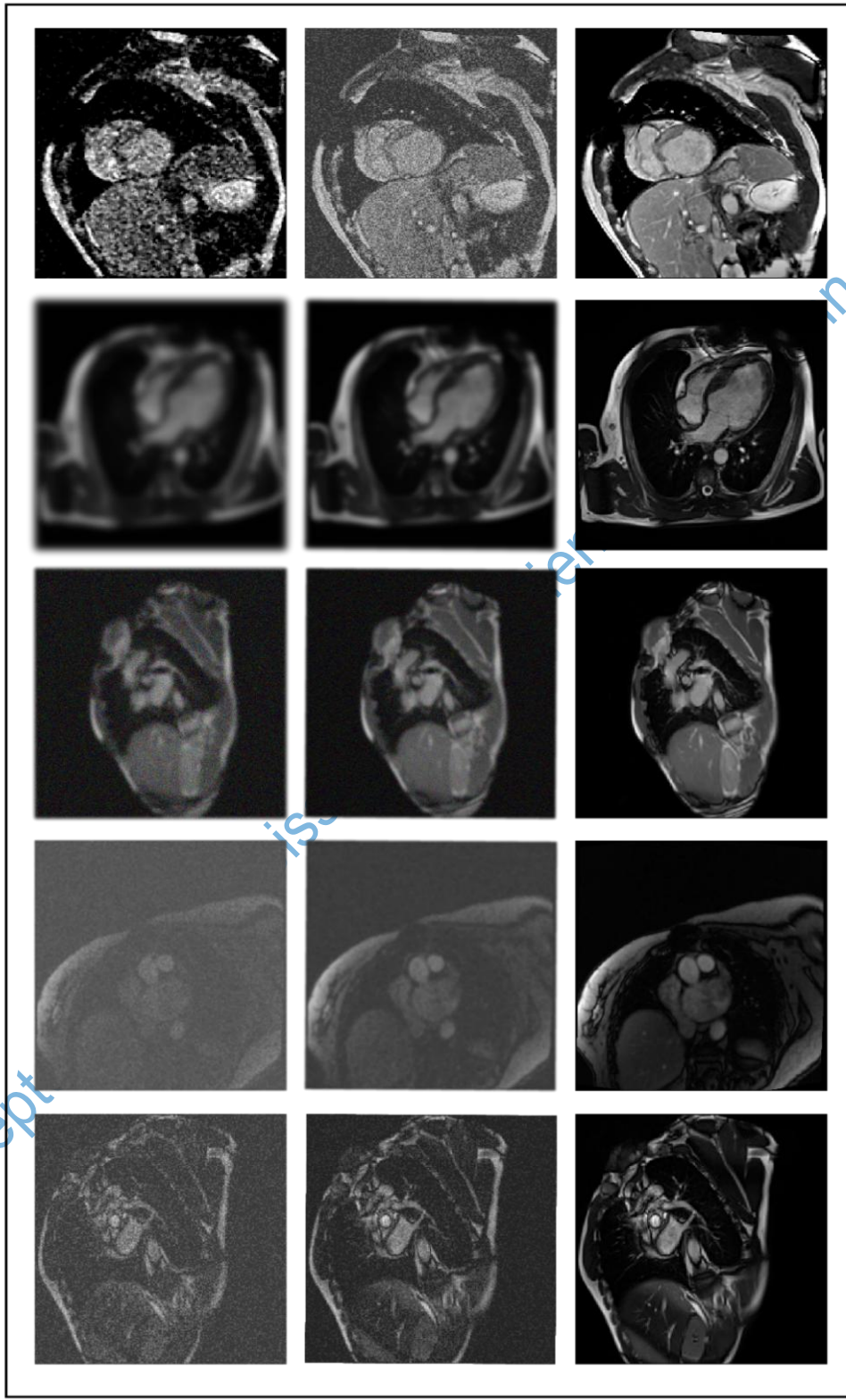

Fig. 6: CSR-GAN Testing Results

## 5. Conclusion

This study focused on GAN-based MR images enhancement with the development of SOUP-GAN and CSR-GAN models. The models were specifically enhanced in order to enhance the output quality of MR images models, enhancing the PSNR and SSIM values considerably for the MR images models. CSR-GAN had a better preservation of structure with a PSNR of 34.6 dB and an SSIM of 0.89, whereas SOUP-GAN was best at producing visually smoother images with a PSNR value of 34.4 dB and SSIM of 0.83. The key improvements introduced for the generators and discriminators were the greater complexity of their architectures, more efficient loss functions, and finally, improved training procedures. Early stopping technique, spectral normalisation, and gradient clipping. These techniques have been implemented in the models to ensure the high stability and image quality of the reconstructed images.

Therefore, the future work might involve applying these models to other modalities such as CT scans and Ultrasound imaging. Extending the current models to provide real-time clinical modalities and their architectures refined through transfer learning methods may provide vistas of more sophisticated medical image processing.

| S# | Article | Techniques used | Evaluation Metrics | Results | Overall Result |
|---|---|---|---|---|---|
| 1 | A new generative adversarial network for medical images super-resolution [14] (2022) | Novel GAN with multi-scale shallow features | PSNR, SSIM | **PSNR:** 30.04 dB (STARE), 30.12 dB (DRIVE) **SSIM:** 0.86 (STARE), 0.87 (DRIVE) **Acc:** 86%, 87% | Outperforming prior SR methods by addressing single-scale and single-step upscaling limitations. |
| 2 | CSR-GAN: Medical Image Super-Resolution Using A Generative Adversarial Network [15](2020) | CSR-GAN with adversarial learning for MR images reconstruction | PSNR, SSIM, loss functions | **PSNR:** 33.2 dB, **SSIM:** 0.85, **Acc:** 85% | Better noise reduction and MR images reconstruction |

| 3 | SOUP-GAN: Super-Resolution MR images using Generative Adversarial Networks [16](2022) | SOUP-GAN optimised for perceptual quality | PSNR, SSIM, perceptual loss | **PNSR:** 0.82, **Acc:** 82% | No fully matched ground-truth for real thick-to-thin slice scenarios, so absolute metrics are limited. |
|---|---|---|---|---|---|
| 5 | Proposed Optimised CSR-GAN & SOUP-GAN | Optimised CSR-GAN & SOUP-GAN with deeper networks, spectral normalisation, and advanced activation | PSNR, SSIM | **PSNR:** 34.6 dB (CSR) 34.4 (SOUP) **SSIM:** 0.89 (CSR), 0.83 (SOUP) **Acc:** 89.4% (CSR), 87.3% (SOUP) | Best balance of detail preservation and noise reduction |

Tab.5: Comparison of Proposed GAN models with State of the Art